\documentclass[12pt]{article}

\usepackage{scicite}
\usepackage{amsmath}

\usepackage{times}

\newcounter{lastnote}

\usepackage{tikz}
\usetikzlibrary{positioning, calc, shapes.geometric, shapes.multipart,
shapes, arrows.meta, arrows,
decorations.markings, external, trees}
\tikzstyle{Arrow} = [
thick,
decoration={
markings,
mark=at position 1 with {
\arrow[thick]{latex}
}
},
shorten >= 3pt, preaction = {decorate}
]

\begin{document}

\baselineskip24pt


\title{A multistate approach for mediation analysis in the presence of semi-competing risks with application in cancer survival disparities}

\author
{\hspace{-0.5in}Linda Valeri$^{1\ast}$, Cecile Proust-Lima$^{2}$, Weijia Fan$^{1}$, \\
\hspace{-0.5in}Jarvis T. Chen$^{3}$,   Helene Jacqmin-Gadda$^{2}$\\
\\
\normalsize{\hspace{-0.5in}$^{1}$Department of Biostatistics, Columbia  University Mailman School of Public Health,}\\
\hspace{-0.5in}\normalsize{722 W 168th St, New York, NY, USA}\\
\normalsize{\hspace{-0.5in}$^{2}$Department of Biostatistics and Epidemiology, Universite de Bordeaux, Talence, France}\\
\normalsize{\hspace{-0.5in}$^{3}$Department of Social and Behavioral Sciences, Harvard T.H. Chan School of Public Health, Boston, MA, USA}\\
\\
\normalsize{\hspace{-0.5in}$^\ast$To whom correspondence should be addressed; E-mail:  lv2424@columbia.edu.}
}

\date{}

\maketitle
\newpage

\vspace{-1in}

\section{Abstract}
We propose a novel methodology to quantify the effect of stochastic interventions on non-terminal time-to-events  that lie on the pathway between an exposure and a terminal time-to-event outcome. Investigating these effects is particularly important in health disparities research when we seek to quantify inequities in timely delivery of treatment and its impact on patients’ survival time. Current approaches fail to account for semi-competing risks arising in this setting.
Under the potential outcome framework, we define and provide identifiability conditions for causal estimands for stochastic direct and indirect effects. Causal contrasts are estimated in continuous time within a multistate modeling framework and analytic formulae for the estimators of the causal contrasts are developed. We show via simulations that ignoring censoring in mediator and/or outcome time-to-event processes, or ignoring competing risks may give misleading results. This work demonstrates that rigorous definition of the direct and indirect effects and joint estimation of the outcome and mediator time-to-event distributions in the presence of semi-competing risks are crucial for valid investigation of mechanisms in continuous time. We employ this novel methodology to investigate the role of delaying treatment uptake in explaining racial disparities in cancer survival in a cohort study of colon cancer patients.

\section{Introduction}
\label{sec1}

Mediation analysis is a popular approach to decompose the total effect of an exposure on an outcome into direct and indirect effects through an intermediate factor, called a mediator (Greenland and Robins, 1992). Literature in this field is fast growing, now accommodating a wealth of designs, variable dimensions and types (VanderWeele, 2015). Along with the estimation of ``natural" effects, other effects that involve interventions on mediators might be of interest. Specifically, controlled direct effects represent the effect of the exposure on the outcome while the mediator is further fixed to a specific level or an intervention on its distribution is sought (i.e., stochastic interventions) (Vansteeland and Daniel, 2017; Muñoz and van der Laan, 2012).\newline
Recently, the causal inference community has drawn attention to the study of causal effects in the presence of competing or semi-competing risks. Competing risks arise when a terminal failure time outcome competes with another terminal event (Andersen et al., 2012). Semi-competing events arise when the outcome of interest is a non-terminal event that competes with a terminal time-to-event (e.g. death) which can occur before or after the outcome of interest (Fine, Jiang and Chappell, 2001). The consequence of both competing and semi-competing events is to render non observable, effectively undefined, the primary outcome of interest when the competing event occurs first, a phenomenon that is also referred to as ``truncation" (Zhang and Rubin, 2003). This poses challenges to formalizing and estimating causal contrast for total effects as well as direct and indirect effects. In particular, excluding from the analysis individuals for whom the competing event occurs first and the outcome is not observed, has been shown to lead to selection bias (Tchetgen et al., 2012). \newline
We consider the setting in which interest lies in estimating the effect of interventions on a non-terminal time-to-event mediator in the pathway between the exposure and a terminal time-to-event outcome. 
The present work is motivated by the study of the role of disparities in access to care as determinants of racial disparities in survival among cancer patients. Recently, causal inference approaches have been proposed to investigate determinants of such disparities (VanderWeele and Robinson, 2014; Valeri et al., 2016, Devick et al., 2020). We here wish to investigate the role of waiting time to surgery in explaining racial disparities survival among colon cancer (CC) patients adopting a causal inference approach. Specifically, we wish to estimate the extent to which racial disparities in survival would remain had the distribution of the intermediate time-to-event, time to surgery, in the least advantaged population been equalized to that observed in the most advantaged population. We address this question in the Cancer Care Outcomes Research and Surveillance Consortium (CanCORS) patient population. A  challenge in quantifying the impact of this intervention arises from the fact that the time to surgery is partially unobserved, as some patients died prior to receiving treatment. Exploratory analyses suggest that Black colon cancer patients are more likely to die before receiving treatment than White patients. Furthermore, among the ones who survive, Blacks display a longer waiting time to surgery. We here set to address this question overcoming the challenge of semi-competing risks. \newline
Our work builds upon the substantial body of literature that describes and proposes solutions to the problem of semi-competing and competing risks when estimating total effects, direct and indirect effects.  Recently, solutions to define, identify and estimate total effects in the presence of truncation have been proposed. When the outcome is a time-to-event, Young, Tchetgen, and Hernán (2018) and Stensrud et al. (2020) proposed novel causal estimands for the average treatment effect in the presence of competing risks. Others focused on the study of survivor average causal effects (SACE) for non-survival outcomes in the presence of semi-competing risks (Ding et al., 2011; Wang, Zhou and Richardson, 2017; Comment et al., 2019). Yet, 
 the problem of semi-competing risks has received relatively little attention when interest lies in estimating direct and indirect effect on a failure time outcome.
Proposals for novel causal contrasts and g-methods for mediation analysis in the context of semi-competing events have been proposed for the setting in which the outcome is a failure time and the mediator is time-dependent (Aalen et al., 2020; Zheng and Van der Laan, 2017; Lin et al., 2017; and Tai et al., 2020). 
Although tremendous progress has been made to advance the formulation of causal effects in the presence of semi-competing events, unsolved questions remain. First, thus far no causal contrasts have been proposed in the context of mediation analysis when both mediator and outcome are times-to-event. Second, causal mediation contrasts put forth by the most recent contributions where the outcome is a time-to-event and the mediator fixed time or longitudinal are defined on the hazard ratio scale. Several authors have shown that regardless of how competing events are accommodated, contrasts of hazards cannot generally be interpreted as causal effects (Hernán, 2010). Third, most commonly mediation analysis proceeds by modeling the mediator and outcome regressions separately. In the presence of non-linearities, this estimation strategy is often affected by a model incompatibility issue that precludes valid effect decomposition. Such issues could be particularly exacerbated when both outcome and mediator are potentially censored. \newline
To address these gaps, we propose interventional analogues to direct and indirect effects to estimate the effect of stochastic interventions on a time-to-event mediator in the presence of semi-competing risks.  We provide conditions that lead to non-parametric identification and propose a novel multistate modeling approach for estimation and inference of the causal effects. We further demonstrate the necessity to properly account for semi-competing risks via simulations and illustrate our methodology in the study of racial disparities in colon cancer survival.

\section{Methods}
\label{sec2}


\subsection{Notation and causal estimands in the presence of semi-competing risks}
Let $A$ denote the exposure variable (race in our context) and $X$ denote cancer stage at diagnosis ($I-IV$), capturing the illness severity when the cancer is diagnosed. Let $T$, $S$ and $K$ denote the time to surgery, the time to death and the time to censoring, respectively.  Censoring time is assumed to be non informative. The observation times for $T$ (a non-terminal event) and $S$ (a terminal event) processes are denoted by $Y^T = min(T,S,K)$ and $Y^S = min(S,K)$ with event observation indicators $\delta^T = I(Y^T=T)$ and $\delta^S = I(Y^S = S)$. We also observe $C$, a vector of baseline covariates that are confounders of the $T$-$S$ relationship. The observed data for individual $i$ is $O_i = (Y_i^T,\delta_i^T,Y_i^S,\delta_i^S ,X_i,A_i,C_{i})$.
\newline
The  directed acyclic graph 
in Figure 1A describes stage and time to treatment as determinants of racial differences in cancer survival. In particular, for patients in each stage, we are interested in understanding the role of racial differences in time to treatment from diagnosis in explaining the racial differences in survival.\newline 
We can define racial disparities (TE, total effect) in survival among colon cancer patients using different scales. For example,  we might be interested in estimating racial disparities with respect to the survival function (eq. 2.1) or we could also consider the restricted mean survival time, restricted at time $r$ (eq. 2.2). 

\begin{small}
\begin{eqnarray}
TE_s=P(S>s|A=1,X,C)-P(S>s|A=0,X,C)
\end{eqnarray}
\begin{eqnarray}
TE_r=E(min(S,r)|A=1,X,C)-E(min(S,r)|A=0,X,C)
\end{eqnarray}
\end{small}

We are interested in  a fixed or stochastic intervention $g$ on the mediator. $G(\cdot)$ denotes a random draw from an arbitrary distribution in a hypothetical population. This distribution could be completely synthetic and of arbitrary choice of the investigator, or learned from the data. In the latter case,  $g=G(T_{x,c})(A=a)=G(T_{x,c})(a)$ denotes a random draw from the time to treatment distribution observed among  individuals with $A=a$, $X=x$ and $C=c$.\\
We can consider survival differences across exposure under an intervention that fixes the mediator distribution in both exposure groups:

\begin{small}
\begin{eqnarray}
SDE_s=P(S_{g}>s|A=1,X,C)-P(S_{g}>s|A=0,X,C)
\end{eqnarray}
\end{small}


We can interpret (eq. 2.3) as the residual difference in the probability of surviving after time $s$ between White and Black patients had the distribution of time to treatment $g$ been the same across the two exposure groups. Setting $g=G(T_{x,c})(0)$ we formalize a stochastic intervention for which the time to treatment $T$ of all individuals in group $A=1$ with certain stage $x$ and baseline covariates $c$ is randomly assigned as sampled from the distribution of $T$ in the group of individuals with $A=0$, $X=x$, and $C=c$. We refer to this causal contrast as ``stochastic direct effect" ($SDE_s$) because this causal effect fixes the intermediate time-to-event distribution to be the same in both exposure groups. \\
We can also consider causal contrasts within an exposure group considering fixed or stochastic interventions, $g$ and $g^{'}$ on the intermediate time-to-event:

\begin{small}
\begin{eqnarray}
SIE_s=P(S_{g}>s|A=a,X,C)-P(S_{g^{'}}>s|A=a,X,C)
\end{eqnarray}
\end{small}

We can interpret (eq. 2.4) as the change in the probability of surviving after time $s$ for subjects in exposure group $A=a$ for a change in the time to treatment level or distribution from $g$ to $g^{'}$. For example setting $g=G(T_{x,c})(0)$ and $g^{'}=G(T_{x,c})(1)$ indicates a change in the time to treatment distribution from what observed in  group $A=1$ (Black patients in our case) to what observed among group $A=0$ (White patients).  We refer to this causal contrast as ``stochastic indirect effect" ($SIE_s$) because, for this effect to be different from zero, race has to be associated with the intermediate time-to-event which in turn needs to be causally related to the survival time.\newline
In what follows we focus on the causal estimand in (eq. 2.3), the ``residual disparity" or more generally called the ``stochastic direct effect" ($SDE_s$) but the results  apply also to the ``stochastic indirect effect".
Note that for these causal estimands we are not requiring the time to treatment to be always observed. We allow for censoring (under a random mechanism conditional on observed covariates)  and for the terminal event to occur prior to the intermediate event.\newline

\subsection{Identifiability conditions}
Several assumptions must be met to identify these causal contrasts involving the potential outcome $S_{g}$.\\ 
{\it Assumption 1.} Consistency of potential outcomes.\newline
Let $S_i|T_i=t$ denote the survival time of individual $i$ given that this subject is observed to receive treatment at time $t$ and let $S_{i,g=t}$ denote the potential survival time for subject $i$ had we intervened setting the treatment time equal to $t$. For each subject $i$ and each level of time to treatment $T=t$ we assume

\begin{small}
\[
S_{i,g=t}=S_i|T_i=t 
\]
\end{small}

That is for each $i$ and each $t$, the survival in a world where we intervene setting the time to treatment to a specific value $t$ (via a fixed or stochastic intervention) is the same as the survival in the real world where we observe a time to treatment equal to $t$.\\
{\it Assumption 2.} Conditional exchangeability (no unmeasured confounding).
The observed time-to-treatment  assignment does not depend on the potential outcomes after accounting for the set of measured covariates $A$, $X$, and $C$. 

\begin{small}
\[
S_{g} \perp T | A, C,X 
\]
\end{small}

{\it Assumption 3.} Non-informative censoring of event times. The vector of censoring times $K$ is conditionally independent of all potential event times (which implies that the observed censoring time is conditionally independent of all potential event times).

\begin{small}
\[
S_g \perp K | A, C,X 
\]
\end{small}

{\it Assumption 4.} We assume that positivity holds. That is for each covariates pattern $(A,X,C)$ that has positive probability in the data, such that the joint density $f(A,X,C) > 0$ is positive and the probability of time to treatment (fixed or stochastic) intervention $g=t$ is positive (i.e. $ Pr(g=t | A,X,C) > 0$) with probability 1.\\
Under these assumptions, the g-formula of non parametric identification for the residual disparity  (eq. 2.3) where $g=G(T_{x,c})(0)$ is:

\begin{small}
\begin{equation}
\begin{split}
&P(S_{G(T_{x,c})(0)}>s|A=1,X,C)-P(S_{G(T_{x,c})(0)}>s|A=0,X,C)\\
& =\int_t P(S_{t}>s|G(T_{x,c})(0)=t,A=1,X,C)f_{G(T_{x,c})(0)}(t|A=1,X,C)dt \\
& \hspace{0.1in} - \int_t 
P(S_{t}>s|G(T_{x,c})(0)=t,A=0,X,C)f_{G(T_{x,c})(0)}(t|A=0,X,C)dt \\
&= \int_t P(S_{t}>s|G(T_{x,c})(0)=t,A=1,X,C)f_{G(T_{x,c})(0)}(t|A=1,X,C)dt-P(S>s|A=0,X,C) \\
&= \int_t P(S_{t}>s|T=t,A=1,X,C)f(t|A=0,X,C)dt-P(S>s|A=0,X,C)\\
&= \int_t\left\{P(S>s|T=t,A=1,X,C)-P(S>s|T=t,A=0,X,C)\right\}f(t|A=0,X,C)dt \\
\end{split}
\end{equation}
\end{small}

where, the first equality is due to the assumption of conditional exchangeability, the second equality solves the integral for the $A=0$ group, the third equality uses the definition of stochastic intervention that fixes the random variable $T$ to $t$, and the fourth equality uses consistency and re-expands the integral for the $A=0$ group.\newline

\subsection{Estimators under a multistate model representation}

The multistate model (Figure 1B) is a natural framework to operationalize the nonparametric estimator derived in the previous section. Our setting involves the following two scenarios: (1)  the individual  transits from diagnosed to treated, and then from treated to death or (2)  the individual transits from diagnosed directly to death and time to treatment is censored at time of death. In this context our intervention aims at controlling the intensity of the transition from diagnosed to treated only.

We define $\alpha_{01}(t|A,X,C)$, $\alpha_{02}(t|A,X,C)$ and $\alpha_{12}(t|t^{'},A,X,C)$ as the instantaneous hazard of transiting from diagnosed to treated, from diagnosed to death, and from treated to death, respectively. $\Lambda_{01}(s|A,X,C)$, $\Lambda_{02}(s|A,X,C)$, $\Lambda_{12}(s|T,A,X,C)$ are the corresponding cumulative transition intensity functions.
The survival function is the sum of the probability of being  alive and not treated, $P_{00}(s|A,X,C)$, and the probability of being alive and treated, $P_{01}(s|A,X,C)$:

\begin{small}
\[
P(S>s|A,X,C)=P_{00}(s|A,X,C)+P_{01}(s|A,X,C) 
\]
\end{small}

where the probability of being alive and untreated at time $s$ is:

\begin{small}
\begin{eqnarray}
P_{00}(s|A,X,C)& = & e^{-\Lambda_{01}(s|A,X,C)-\Lambda_{02}(s|A,X,C)}
\end{eqnarray}
\end{small}

the probability of being alive and treated at time $s$ is:

\begin{small}
\begin{eqnarray}
P_{01}(s|A,X,C)& = &
 \int^{s}_0 e^{-\Lambda_{01}(u|A,X,C)-\Lambda_{02}(u|A,X,C)}\alpha_{01}(u)e^{-\Lambda_{12}(s|T=u,A,X,C)+\Lambda_{12}(u|T=u,A,X,C)}du
\end{eqnarray}
\end{small}

Let $g=G(T_{x,c})(0)$ be a random draw from the time to treatment distribution in the group $A=0$. The probability of being alive at time $s$ 
and untreated at time $s$ according to the time to treatment distribution observed in the $A=0$ subgroup, $P^{g}_{00}(s|A,X,C)$, and the probability to be alive at time $s$ 
and treated at time $s$ according to the time to treatment distribution observed in the $A=0$ subgroup, $P^{g}_{01}(s|A,X,C)$, are given by:

\begin{small}
\begin{eqnarray}
P^{g}_{00}(s|A,X,C)= e^{-\Lambda_{01}(s|A=0,X,C)-\Lambda_{02}(s|A,X,C)}
\end{eqnarray}
\end{small}

and 

\begin{small}
\begin{eqnarray*}
P^{g}_{01}(s|A,X,C)& = &
\int^{s}_0 e^{-\Lambda_{01}(u|A=0,X,C)-\Lambda_{02}(u|A,X,C)}\alpha_{01}(u|A=0,X,X_0) \\
\end{eqnarray*}
\begin{eqnarray}
& & e^{-\Lambda_{12}(s|T=u,A,X,C)+\Lambda_{12}(u|T=u,A,X,C)}du
\end{eqnarray}
\end{small}

Our estimator for the stochastic direct effect defined in (2.3) can be obtained as the combination of the four probabilities given in equations (2.6)-(2.9):

\begin{small}
\begin{eqnarray*}
SDE_s=P(S_{g}>s|A=1,X,C)-P(S_{g}>s|A=0,X,C)
\end{eqnarray*}

\begin{eqnarray}
=  \{P^{g}_{00}(s|A=1,X,C)+P^{g}_{01}(s|A=1,X,C)\} -\{P_{00}(s|A=0,X,C)+P_{01}(s|A=0,X,C)\}.
\end{eqnarray}
\end{small}

We can interpret this estimator in equation (2.10) as the average difference in the global survival of a hypothetical population of Black patients having the time to treatment distribution observed among White patients who survived to receive treatment and the global survival observed among White patients. The 1st term is the probability of being alive and untreated at $s$ for a Black patient with the treatment transition hazard (from untreated to treated) of a White patient and the  death transition hazard of a Black patient. The second term is the probability of being alive and treated at $s$ for a patient with the treatment transition hazard of a White patient and the death transition hazard of a Black patient given the intervention on the time to treatment. 
The third term is the probability to be alive and untreated at $s$ for a  White patient.
The fourth term is the probability to be alive and treated at $s$ for a White patient. \\

Similarly, it can be shown that the estimator for $SIE_s$ introduced in (2.4) is given by:\newline

\begin{small} 
\begin{eqnarray*} 
SIE_s &= P(S_{G(T_{x,c})(1)}>s|A=1,X,C)-P(S_{G(T_{x,c})(0)}>s|A=1,X,C)\\
\end{eqnarray*}
\begin{eqnarray*}
&=\displaystyle\int_0^{s}P(S>s|t,A=1,C,X)f(t|A=1,C,X)dt-\displaystyle\int_0^{s}P(S>s|t,A=1,C,X)f(t|A=0,C,X)dt\\
\end{eqnarray*}
\begin{eqnarray*}
&=\Big\{P_{00}(s|A=1,X,C)+P_{01}(s|A=1,X,C)\Big\}-\Big\{P^{g^{'}}_{00}(s|A=1,X,C)+P^{g^{'}}_{01}(s|A=1,X,C)\Big\} \\
\end{eqnarray*}
\begin{eqnarray*}
&=  \Big\{e^{-\Lambda_{01}(s|A=1,X,C)-\Lambda_{02}(s|A=1,X,C)}+\\
&+\int_0^s e^{-\Lambda_{01}(t|A=1,X,C)-\Lambda_{02}(t|A=1,X,C)}\alpha_{01}(t|A=1,X,C)e^{-\Lambda_{12}(s|t,A=1,X,C)+\Lambda_{12}(t|t,A=1,X,C)} dt \Big\}\\
\end{eqnarray*}
\begin{eqnarray*}
&- \Big\{e^{-\Lambda_{01}(s|A=0,X,C)-\Lambda_{02}(s|A=1,X,C)}\\
\end{eqnarray*}
\begin{eqnarray}
&+  \int_0^s e^{-\Lambda_{01}(t|A=0,X,C)-\Lambda_{02}(t|A=1,X,C)}\alpha_{01}(t|A=0,X,C)e^{-\Lambda_{12}(s|t,A=1,X,C)+\Lambda_{12}(t|t,A=1,X,C)} dt \Big\}. 
\end{eqnarray}
\end{small}

\subsection{Semi-parametric Estimators}

We propose a semi-parametric approach to estimate the causal contrasts defined above.
In particular, we specify a semi-parametric proportional intensity model for the hazard of transitioning from diagnosed to treated ($\alpha_{01}(t|A,X,C)$), from diagnosed to death ($\alpha_{02}(t|A,X,C)$) and from treated in $t^{'}$ to death in $t$ ($\alpha_{12}(t|t^{'},A,X,C)$). 

\begin{small}
\begin{eqnarray}
\alpha_{01}(t|A,X,C)
= \alpha^{0}_{01}(t)e^{\beta_1A+\beta_2 X +\beta^{'}_3C}
\end{eqnarray}

\begin{eqnarray}
\alpha_{02}(t|A,X,C) 
= \alpha^{0}_{02}(t)e^{\gamma_1A+\gamma_2 X +\gamma^{'}_3C}
\end{eqnarray}

\begin{eqnarray}
\alpha_{12}(t|t^{'},A,X,C) 
= \alpha^{0}_{12}(t)e^{\delta_1A+\delta_2t^{'}+\delta_3A*t^{'}+\delta_4 X +  \delta^{'}_5C} \hspace{0.2in} \mbox{for} ~~ t^{'} < t 
\end{eqnarray}
\end{small}
\nonumber

where baseline hazards are non-parametrically estimated. We consider here linear relationships for simplicity but the methodology accommodates more flexible formulations on the covariates. \newline
Recalling that the observation times for $T$ (the non-terminal event) and $S$ (the terminal event) processes are denoted by $Y^T = min(T,S,K)$ and $Y^S = min(S,K)$, and that event observation indicators are given by $\delta^T = I(Y^T=T)$ and $\delta^S = I(Y^S = S)$, the observed data likelihood contribution for person $i$ defined by these hazards is:

\begin{small}
\[
\mathcal{L}_i=\left[e^{-\Lambda_{01}(Y_i^T)-\Lambda_{02}(Y_i^T)} \alpha_{01}(Y_i^T)e^{-\Lambda_{12}(Y_i^S)+\Lambda_{12}(Y_i^T)} \alpha_{12}(Y_i^S|Y_i^T)^{\delta_i^{S}} \right]^{\delta_i^{T}} 
\left[e^{-\Lambda_{01}(Y_i^S)-\Lambda_{02}(Y_i^S)} \alpha_{02}(Y_i^S)^{\delta_i^{S}}\right]^{(1-\delta_i^{T})}.
\]
\end{small}

The estimation procedure for the $SDE_s$ and the $SIE_s$ can then seamlessly proceed by fitting the multistate model, estimating the cumulative hazard via the Breslow method (Lin, 2007) and solving the integrals that are involved in the estimator by computing $\alpha$ and $\Lambda$ using rectangular (or Simpson) method (Atkinson and Kendall, 1989) along the time interval $(0,s)$.\newline 
The estimation procedure for the causal quantities can be summarized in three steps: 
\begin{itemize}
    \item[(1)] estimate the multistate regression model coefficients. We here decided to use the R package {\it mstate} (de Wreede et al., 2011) for this purpose but other alternatives can be considered (Jackson, 2011). 
     \item[(2)]predict baseline hazards, hazards, and cumulative hazards for each transition for observed times in the range $t \in (0,s]$, in new data for each racial/ethnic group and time to treatment $t \in (0,s]$. We used the function {\it msfit()} in this step.
     \item[(3)]calculate the causal effects of interest by plugging in the estimated hazards and cumulative hazards.\newline
\end{itemize}  
Inference can be conducted via bootstrapping.
R code for the procedure can be found on the GitHub page of the corresponding author (add url).

\section{Simulation Study}
We conducted an extensive simulation study to evaluate the performance of our proposed method (``multi-state") and to compare it with two other commonly used methods when analyzing semi-competing risk data (``exclude $T>S$" and ``censor $T>S$"). For our proposed method, we include all subjects. The ``exclude $T>S$" approach excludes from the analytic sample  subjects who experienced the terminal event but not the non-terminal event, such that  $T>S$. The ``censor $T>S$" approach includes all subjects in the analysis, treating all subjects who experienced the terminal event but not the non-terminal event as being censored, with the parameters referring to transition from diagnosed to death set to zero. We examined 3 semi-competing risks situations, considering no semi-competing risks (i.e., everyone has either the intermediate event or is censored for $T$ and $S$), 10\% and 40\% semi-competing risks (10\% or 40\% subjects experience the terminal event without experiencing the intermediate event). For each semi-competing risk scenario, we considered 4 scenarios for total effect ($TE_s$), and stochastic direct effect ($SDE_s$) and stochastic indirect effect ($SIE_s$) all generated under a proportional intensity multistate model (2.12)-(2.14), allowing for either a linear or partially linear model (exposure-mediator interaction) specification. In scenarios 1 and 2 we set $SIE_s\neq 0$ and $SDE_s\neq0$, and specify the model for the hazard of transitioning from the intermediate non-terminal event to the terminal time-to-event outcome without (scenario 1) and with  interaction (scenario 2) between the intermediate time-to-event and exposure. In scenario 3 we specify $SIE_s\neq 0$ and $SDE_s=0$ and in scenario 4 we set $SIE_s= 0$ and $SDE_s\neq0$ . \newline
We simulated 100 datasets with sample size of 2000, and we calculated quantities of interest for time points starting from baseline to $s=24$ months follow-up by increments of 0.5 months. Standard errors of the estimates were computed by bootstrap with 100 bootstrap samples. Performance of each method under each scenario was assessed for bias, confidence interval (CI) coverage probability, mean squared error (MSE), and type I error rate (null cases) for effect estimates at 24 months. Full details of the simulation strategy are given in Appendix A of the Supplementary Material available at
\textit{Biostatistics} online.

\subsection{Results}
Figure 2 compares the performance of different methods for scenario 1 with 10\% and 40\% semi-competing risks. Figures S1-S3 in Appendix B of the Supplementary Material show results for the other simulation scenarios.  \newline
For all four effect type scenarios and all levels of semi-competing risks, the proposed multistate approach provides unbiased estimates, CI coverage probability close to 95\% and lowest MSE compared to the alternative approaches for all three effects of interest.  The  approaches ``exclude $T>S$ and ``censor $T>S$" produce biased estimates and suboptimal coverage for $SDE_s$ and $SIE_s$ even when semi-competing risks are low (10\%) (Figures 2A, 2C and S1). As expected, for the estimation of $SDE_s$ under the scenario in which $SDE_s=0$ and for the estimation of $SIE_s$ under the scenario in which $SIE_s=0$ (Figures S2 and S3) all approaches are unbiased. For all other scenarios, the estimates for $SDE_s$ and $SIE_s$ may be severely biased and confidence intervals coverage probability are not satisfying, when these alternative approaches are adopted. \newline
 
The simulation results underscore that severe bias will arise when semi-competing risks are ignored and bias will increase as the semi-competing risks increase (Figures 2B and 2D) and in the presence of non linearities, such as an exposure-mediator interaction (Figure S2). Type I errors are invalid for the test of the $SIE_s$ when semi-competing risks are high and the ``exclude $T>S$" method is adopted (Figure S6). The simulation study demonstrates that adopting the multistate approach to mediation analysis that we propose appropriately addresses semi-competing risks under the identifiability assumptions discussed in section 2.2 and correct model specification. \newline

\section{Application}
We obtained data from the National Cancer Institute’s CanCORS Consortium, which contains detailed information from cancer patients and physicians (Ayanian et al., 2004). The consortium collected data on colorectal cancer (CRC) cases in multiple regions and health care delivery systems across the U.S. The study population consisted of non-Hispanic White and non-Hispanic Black patients enrolled between 2003 and 2005. To be eligible for enrollment as a CRC case, the study required patients to be at least 21 years old and with newly diagnosed invasive adenocarcinoma of the colon or rectum. The CanCORS Consortium oversampled minority groups at several of the sites. Of the ten CanCORS centers that enrolled patients with CRC, 8 centers collected survival data, including: Henry Ford Health System (HFHS), Kaiser Permanente Hawaii (KPHI), Kaiser Permanente Northwest (KPNW), Northern California Cancer Center (NCCC), State of Alabama (UAB), Los Angeles County (UCLA), North Carolina (UNC), and Veterans Health Administration (VA). 
CanCORS collected stage at diagnosis information via medical record abstraction or from cancer registries, categorized as stage I-IV according to the American Joint Committee on Cancer (AJCC) staging criteria (Edge et al., 2010). For the purpose of the analysis, we selected patients diagnosed with colon cancer (CC), and conducted analyses stratified by stage. We calculated time from diagnosis to treatment from the corresponding dates reported in CanCORS data for treament. We focused on surgery as the treatment of interest since report of dates of other treatments (chemotherapy and radiotherapy) was sparse in this sample and surgery is the first line of therapy for most patients diagnosed with colon cancer across stages (Lawler et al., 2020; Libutti et al., 2019).

\subsection{Modeling strategy}
For our CanCORS data analysis, we used survival time in months since diagnosis. We considered age at diagnosis (age $<50$, $50-65$, and $>65$ years), sex (female versus male), income level ($<\$40k$, $\$40-80k$, $> \$80k$; with the middle group category being the reference), and medical center to be potential confounders of the time to surgery-survival relationship. Table S1 provides descriptive statistics of the sample of CC patients by racial-ethnic group. We inspected the cumulative incidence curves for waiting time to surgery and survival time (Figure 3) by racial ethnic group.
We fit a multistate Cox proportional hazard regression for the transitions from diagnosis to surgery, from diagnosis to death and from surgery to death. Motivated by Valeri et al. (2016), we conducted backwards model selection allowing for race by income/gender/age,  time to surgery by race, and time to surgery by income interactions. We also considered quadratic effects of time to surgery in the transition to death. We performed analyses using the best fitting models for the three transitions.  In CanCORS, we estimated the Black-White disparity in 5-year (60 month) survival probability prior to intervention on the time to surgery distribution ($TE_{60}$). Further, we estimated the residual difference in 5-year (60 month) survival probability  after a hypothetical intervention on the distribution of waiting time to surgery of the Black patients to match the distribution  waiting time to surgery observed among the White patients ($SDE_{60}$) using our multistate modeling approach. We considered our approach along with the two alternative estimation procedures that ignore semi-competing risks, namely ``exclude $T>S$" and ``censor $T>S$". We calculated 95\% confidence intervals for the disparity ($TE_{60}$), the residual disparity under intervention ($SDE_{60}$) and the proportion of the disparity eliminated by the intervention ($PE=\frac{TE_{60}-SDE_{60}}{TE_{60}}$) using the bootstrap with 100 bootstrap samples.

\subsection{Results}
From descriptives in Table S1 and cumulative incidence curves in Figure 3 we note that Black patients had longer waiting time to surgery and slightly shorter survival times on average compared to White patients. We found also that 6.2\% of the patients died prior to receiving treatment, indicating a semi-competing risk problem, albeit small. At diagnosis Black patients were younger and presented higher stages than White patients. We observed income and gender differences as well.\\
We report results of analyses for stage II patients (n=283, comprising 23.2\% of the sample) in the main manuscript. Results for other stages can be found in the Appendix B of the Supplementary Material (Figures S7-S9).\\
Table 1 displays the output of the multistate modeling analyses. Adjusting for gender, age, income and center, there is suggestive evidence that stage II Black patients experienced a reduction in the hazard of surgery compared to White patients (HR=0.75, 95\% confidence interval=0.55 – 1.03) and an increase in hazard of dying prior to receiving surgery (HR=3.26, 95\% confidence interval=0.51 – 20.0). Considering the transition from treatment to death status modeled without adjustment for the mediator, we found evidence of racial-ethnic disparities at the intersection of income category. Black patients within the middle income group  experienced more than three times the hazard of dying compared to the White patients within the middle income group (HR=3.71, 95\% confidence interval=2.19 – 6.30). When adjustment is made for the mediator, the racial coefficient for this income group is reduced (HR=1.39, 95\% confidence interval=0.82 – 2.36).  In addition to race-income interaction, we found suggestive evidence of non-linear effect of time to treatment, whereby the hazard of death increased with longer waiting time to treatment, with this effect reducing over time. \\
Figure 4 displays the estimated racial disparity for stage II, middle income CC patients on the survival probability difference scale  and the residual disparity on the survival probability difference scale had the hazard of transitioning from diagnosis to treatment been the same between the two racial-ethnic groups. Results are shown for the three modeling strategies. Table 2 shows the effect estimates for the difference of 5-year survival probability between the two racial ethnic groups before ($TE_{60}$) and after ($SDE_{60}$) the intervention on the time to treatment distribution. Using the proposed multistate approach, we estimate a Black-White disparity in 5-year survival probability of $TE_{60}=-0.29$, 95\% confidence interval=(-0.51, -0.05) and if we implemented an intervention so that Black and White patients displayed the same distribution for time to surgery, the residual disparity after the intervention would be $SDE_{60}=-0.26$, 95\% confidence interval=(-0.47, -0.04). These results indicate that 8\% of the racial-ethnic disparity would be eliminated by such an intervention. The alternative approach selecting only patients who were treated, leads to a slightly reduced estimated disparity of $TE_{60}=-0.25$, 95\% confidence interval=(-0.49, 0.00) and no change when considering the intervention ($SDE_{60}=-0.25$, 95\% confidence interval=(-0.45, 0.00)). Similar results are obtained from the analysis that considers the semi-competing event as a censoring event. Results for other stages indicate similarly weak impact of the intervention in our sample (Figures S7-S9). Our application shows that ignoring semi-competing risks, even when the contribution of semi-competing risks is small (as in these data), may lead to different estimates of the disparities and the impact of the intervention.

\section{Discussion}

We have protentised a novel approach for mediation analysis when mediator and outcome are times-to-event. To our knowledge, this is the first method presented in the causal inference literature to estimate randomized interventional analogues of direct and indirect effects when the mediator is a time-to-event and semi-competing risks are present. This method allows for joint modeling of outcome and mediator through the multistate model formulation. Our extension of causal mediation methodology that allows for semi-competing risks and interventions on intermediate times-to-event is important for many applications in public health. For instance, this approach could be applied to quantify the change in racial disparities in survival for patients affected by any health condition, say COVID-19, after an intervention on timing of care, say admission to ICU. In other epidemiological setting across the life-course, this approach could be applied to evaluate the role of intermediate time-to-events, in early or mid-life, such as cardiovascular events, in explaining disparities or exposure effects on late life events, such as dementia. \newline
Our simulation study shows our proposed multistate approach performs better than current methods that ignore semi-competing risks. In the presence of such phenomenon, we thus advise to use our approach over other methods.\newline
Applying this novel methodology to the CanCORS cohort, we found a racial disparity in colon cancer survival for middle income stage II patients and a negative association of race and time to surgery. If the time to surgery distribution were fixed for both groups to what observed in the White population, the disparity would be reduced by about 8\%, suggesting that interventions improving timeliness of treatment could potentially reduce slightly racial disparities in survival.\newline
Our approach recovers the counterfactual survival curve in the group of Black patients if their time to treatment distribution were fixed to what was observed among the White patients who survived long enough to receive treatment. This causal interpretation is weaker than the one put forth by natural direct and indirect effects. The identification of natural effects requires stronger assumptions than the ones we have discussed for stochastic direct and indirect effects. We would need to assume that no confounders of the mediator outcome relationship are affected by the exposure and that there are no confounders of the exposure-mediator relationship. Moreover, natural effects in the presence of semi-competing risks require the identification of the counterfactual distribution of the time-to-event mediator when death could be treated either as a conditioning or a censoring event (Young et al., 2018). The extension to natural effects in the presence of competing risks as well as the setting of multiple time-to-event mediators is a direction for future work. \newline
Some limitations of our method are due to the reliance on no unmeasured confounding assumptions and on the correct specification of the multistate model transition probabilities as a function of the exposure, confounders and mediator.  In our data application, our results are limited by several factors such as potential residual confounding by socio-economic factors beyond income, such as education, and measurement error in time to surgery, which was self-reported. The causal effect should be interpreted with caution. The stochastic direct and indirect effects that we propose consider  a change in the waiting time to treatment distribution observed among the White and Black patients who survived.  We might argue whether these distributions are representative of the target population. This  depends on the hazard of the transition from diagnosis to death in the target population, which is a function of both individual and quality of care factors. The generalizability of the racial disparities estimates beyond CanCORS may be limited,  as only research medical centers were involved in the cohort study. Such medical centers typically display higher quality of care. Finally, we here have considered only treatment timing as potential determinants of racial disparities in survival, other aspects of treatment quality should be considered.\newline
In future work, we plan to evaluate the performance of our approach in the presence of measurement error in the time-to-event mediator and to extend the approach to allow for multiple sequential time-to-event mediators and time-dependent confounders. 


\section{Software}

Software in the form of R code, together with a sample
input data set and complete documentation is available on
request from GitHub page: https://github.com/wf2213/multistate.

\section*{Acknowledgments}
This study makes use of data generated by
the CanCORS Consortium.\newline
This work was supported by the National Institute of Mental Health award K01 MH118477.\newline
{\it Conflict of Interest}: None declared.

\section*{References}

    Aalen, O. O., Stensrud, M. J., Didelez, V., Daniel, R., Røysland, K., \& Strohmaier, S. (2020). Time‐dependent mediators in survival analysis: Modeling direct and indirect effects with the additive hazards model. Biometrical Journal, 62(3), 532-549.\newline
    Andersen, P. K., Geskus, R. B., de Witte, T., \& Putter, H. (2012). Competing risks in epidemiology: possibilities and pitfalls. International Journal of Epidemiology, 41(3), 861-870.\newline
    Atkinson, K.E. (1989). An Introduction to Numerical Analysis (2nd ed.). John Wiley \& Sons. ISBN 0-471-50023-2.\newline
    Ayanian, J. Z., Chrischilles, E. A., Wallace, R. B., Fletcher, R. H., Fouad, M. N., Kiefe, C. I., ... \& West, D. W. (2004). Understanding cancer treatment and outcomes: the cancer care outcomes research and surveillance consortium. Journal of Clinical Oncology: official journal of the American Society of Clinical Oncology, 22(15), 2992-2996.\newline
   Comment, L., Mealli, F., Haneuse, S., \& Zigler, C. (2019). Survivor average causal effects for continuous time: a principal stratification approach to causal inference with semicompeting risks. arXiv preprint arXiv:1902.09304.     \newline
      Devick, K. L., Valeri, L., Chen, J., Jara, A., Bind, M. A., \& Coull, B. A. (2020). The role of body mass index at diagnosis of colorectal cancer on Black–White disparities in survival: a density regression mediation approach. Biostatistics. kxaa034,\\ https://doi.org/10.1093/biostatistics/kxaa034\newline
        de  Wreede,  L.  C.,  Fiocco,  M.,   Putter,  H.  (2011).  mstate:  an  R  package  for  the analysis of competing risks and multi-state models. Journal of statistical software, 38(7),1-30.\newline

    Ding, P., Geng, Z., Yan, W., \& Zhou, X. H. (2011). Identifiability and estimation of causal effects by principal stratification with outcomes truncated by death. Journal of the American Statistical Association, 106(496), 1578-1591.\newline
    Edge, S.B., Byrd, D. R., Carducci, M. A., Compton, C. C., Fritz, A. G., \& Greene, F. L. (2010). AJCC cancer staging manual (Vol. 649). S. B. Edge (Ed.). New York: Springer.\newline
    Fine, J.P., Jiang, H.,  Chappell, R. (2001). On semi-competing risks data. Biometrika, 88(4), 907-919.
    Hernán, M.A. (2010). The hazards of hazard ratios. Epidemiology (Cambridge, Mass.), 21(1), 13.
    Jackson,  C.  H.  (2011). Multi-state models for panel data: the msm package for R. Journal of statistical software,38(8), 1-29.\newline
    Lawler, M., Johnston, B., Van Schaeybroeck, S., Salto-Tellez, M., Wilson, R., Dunlop, M., and Johnston, P.G. Chapter 74 – Colorectal Cancer. In: Niederhuber JE, Armitage JO, Dorshow JH, Kastan MB,Tepper JE, eds. Abeloff’s Clinical Oncology. 6th ed. Philadelphia, Pa. Elsevier: 2020.\newline
    Libutti, S.K., Saltz, L.B., Willett, C.G., and Levine, R.A. Ch 62 - Cancer of the Colon. In: DeVita VT, Hellman S,Rosenberg SA, eds. DeVita, Hellman, and Rosenberg’s Cancer: Principles and Practice of Oncology. 11th ed. Philadelphia, Pa: Lippincott-Williams   Wilkins; 2019.\newline
    Lin, D.Y. (2007). On the Breslow estimator. Lifetime data analysis, 13(4), 471-480.
    Lin, S.H., Young, J.G., Logan, R., \& VanderWeele, T.J. (2017). Mediation analysis for a survival outcome with time‐varying exposures, mediators, and confounders. Statistics in medicine, 36(26), 4153-4166.\newline
    Díaz, I.M. and van der Laan, M. (2012). Population intervention causal effects based on stochastic interventions. Biometrics 68.2: 541-549.\newline
Robins, J.M., and Greenland, S. (1992). Identifiability and exchangeability for direct and indirect effects. Epidemiology: 143-155.\newline
    Stensrud, M. J., Young, J. G., Didelez, V., Robins, J. M., \& Hernán, M. A. (2020). Separable Effects for Causal Inference in the Presence of Competing Events. Journal of the American Statistical Association, (just-accepted), 1-9.\newline
    Tai, A. S., Tsai, C. A., \& Lin, S. H. (2020). Survival mediation analysis with the death-truncated mediator: The completeness of the survival mediation parameter. Harvard University Biostatistics Working Paper Series. Working Paper 223.\\
https://biostats.bepress.com/harvardbiostat/paper223 \newline    
 Tchetgen, E. J. T., Glymour, M. M., Shpitser, I., \& Weuve, J. (2012). Rejoinder: to weight or not to weight? On the relation between inverse-probability weighting and principal stratification for truncation by death. Epidemiology, 23(1), 132-137.\newline
    Valeri, L., Chen, J. T., Garcia-Albeniz, X., Krieger, N., VanderWeele, T. J., \& Coull, B. A. (2016). The role of stage at diagnosis in colorectal cancer black–white survival disparities: a counterfactual causal inference approach. Cancer Epidemiology and Prevention Biomarkers, 25(1), 83-89.\newline
    VanderWeele, T.J. (2015). Explanation in causal inference: methods for mediation and interaction. Oxford University Press.\newline
    VanderWeele, T. J., \& Robinson, W. R. (2014). On causal interpretation of race in regressions adjusting for confounding and mediating variables. Epidemiology (Cambridge, Mass.), 25(4), 473-484.\newline
    Vansteelandt, S., \& Daniel, R. M. (2017). Interventional effects for mediation analysis with multiple mediators. Epidemiology (Cambridge, Mass.), 28(2), 258.\newline
    Wang, L., Zhou, X. H., \& Richardson, T. S. (2017). Identification and estimation of causal effects with outcomes truncated by death. Biometrika, 104(3), 597-612.\newline
    Young, J. G., Tchetgen Tchetgen, E. J., \& Hernán, M. A. (2018). The choice to define competing risk events as censoring events and implications for causal inference. arXiv preprint arXiv:1806.06136.\newline
    Zhang,  J.  L.,  \&  Rubin,  D.  B.  (2003).  Estimation of causal effects via principal stratification when some outcomes are truncated by “death”. Journal of Educational and Behavioral Statistics, 28(4), 353-368.
    Zheng, W., \& van der Laan, M. (2017).\newline Longitudinal mediation analysis with time-varying mediators and exposures, with application to survival outcomes. Journal of Causal Inference, 5(2). DOI: https://doi.org/10.1515/jci-2016-0006.

\newpage
\begin{figure}[htbp]
\hspace*{-3.5cm} 
\includegraphics[height=9.5in,width=8.5in]{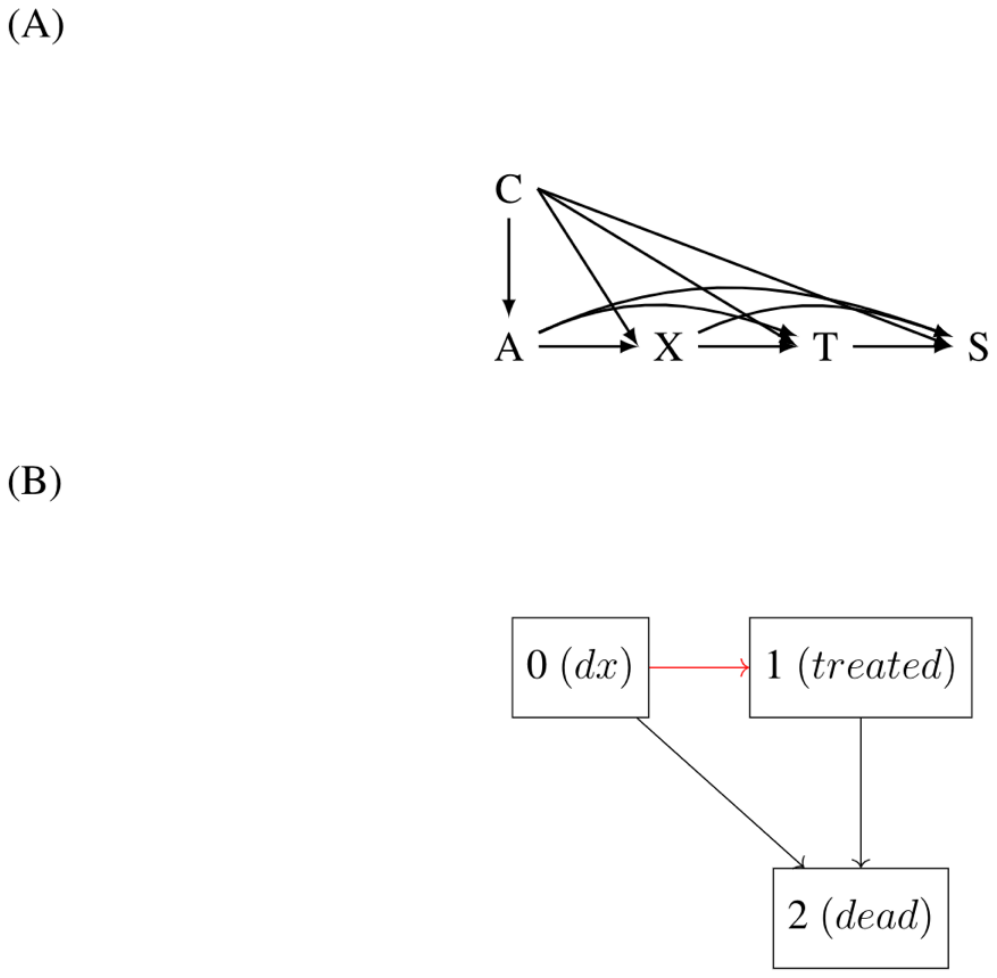}
\vspace{-3in}
\caption{Fig. 1 (A) Directed Acyclic Graph encoding our assumptions on conditional independences among the nodes. (B) Illness-death model representation of our study where individuals can transit from diagnosed (dx) to treated and then to death or can transit directly from diagnosed to death status.}
\end{figure}

\vspace{-1in}

\begin{figure}[htbp]
\hspace*{-4cm} 
\includegraphics[height=9.5in,width=9.5in]{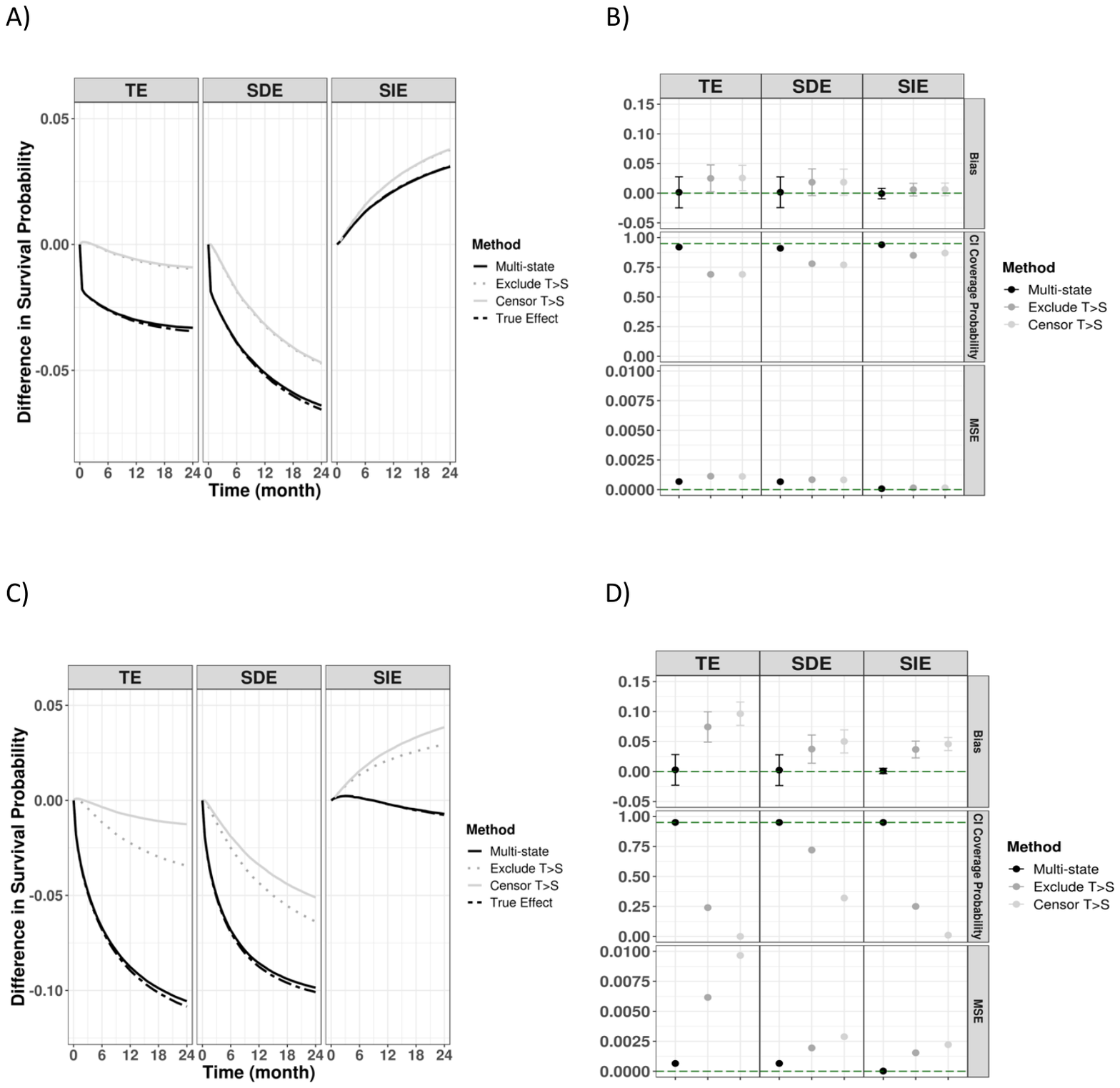}
\vspace{-3in}
\caption{Simulation results for scenario 1 ($SDE_s\neq0$ and $SIE_s\neq0$ no exposure-mediator interaction).\newline
A) and C) Curves of the effects on the survival probability difference scale for 10\% and 40\% competing risks, respectively.\newline
B) and D) Comparison of approaches in the estimation of the effects on differences in the probability of surviving after 24 months in terms of bias, coverage probability and MSE for 10\% and 40\% competing risks, respectively.}

\end{figure}

\setcounter{table}{1}
\begin{table}[!p]
\caption{
\label{Table2} Multistate Cox Proportional Hazard model for transitions diagnosed-treated, diagnosed-death, treated-death in stage II colon cancer patients. 
First column without and second column with adjustment for time to surgery in the third transition. Models are adjusted for: income, age, gender, 
medical center and allow in the third transition for race-income, race-time to surgery interaction, nonlinear effect of time.}

\tabcolsep=4.25pt
\begin{tabular}{@{}lcccccc@{}}
\bf{Predictors}	 &  \bf{Est}&	\bf{CI} & \bf{p}	&   \bf{Est}	& \bf{CI} &	\bf{p}  \\
\bf{Transition diagnosed-treated} &  &	 & 	&   	&  &	\\
Race&	0.75&	0.55 – 1.03	&0.072&	0.75&	0.55 – 1.03	&0.072\\
Age&	0.97&	0.80 – 1.16	&0.713&	0.97&	0.80 – 1.16	&0.713\\
Gender&	1.10&	0.84 – 1.43	&0.492&	1.10&	0.84 – 1.43	&0.492\\
Income $<40K$&	1.07&	0.79 – 1.44&	0.662&	1.07&	0.79 – 1.44&	0.662\\
income $>80K$&	0.85&	0.58 – 1.24&	0.401&	0.85&	0.58 – 1.24&	0.401\\
center [1]&	1.45&	0.89 – 2.36&	0.137&	1.45&	0.89 – 2.36&	0.137\\
center [2]&	2.22&	1.47 – 3.35	&0.001&	2.22&	1.47 – 3.35	&0.001\\
center [3]&	0.76&	0.48 – 1.21	&0.250	&0.76	&0.48 – 1.21	&0.250\\
center [4]&	0.94&	0.62 – 1.42	&0.772	&0.94	&0.62 – 1.42	&0.772\\
center [6]&	1.11&	0.68 – 1.84	&0.672	&1.11	&0.68 – 1.84	&0.672\\
\bf{Transition diagnosed-death}	 &  &	 & 	&   	&  &	\\	
Race&	3.26&	0.51 – 20.4&	0.207&	3.26&	0.51 – 20.4&	0.207\\
Age &	1.25&	0.32 – 4.78&	0.749&	1.25&		0.32 – 4.78&	0.749\\
Income& 	0.97&	0.41 – 2.24&	0.942&	0.97&	0.41 – 2.24&	0.942\\
Gender&	1.84&	0.27 – 12.51&	0.534&	1.84&	0.27 – 12.51&	0.534\\
Center&	1.44&	0.84 – 2.44&	0.178&	1.44&	0.84 – 2.44&	0.178\\
\bf{Transition treated-death}		 &  &	 & 	&   	&  &	\\					
Race& 	3.71&	2.19 – 6.30&	0.029&	1.39&	0.82 – 2.36&	0.780\\
Age &	1.30&	0.63 – 2.66&	0.193&	1.28&	0.62 – 2.65&	0.213\\
Gender &	1.08&	0.39 – 2.99&	0.794&	1.09&	0.39 – 3.03&	0.772\\
Income $<40K$&	2.13&	0.27 – 16.8&	0.039&	2.18&	0.27 – 17.4&	0.035\\
Income  $>80K$&	0.93&	0.93 – 0.93&	0.895&	0.98&	0.98 – 0.98&	0.973\\
Center  [1]&	0.66&	0.20 – 2.15&	0.386&	0.83&	0.08 – 8.39&	0.710\\
Center  [2]&	0.76&	0.30 – 1.94&	0.510&	0.94&	0.36 – 2.44&	0.884\\
Center  [3]&	0.49&	0.22 – 1.11&	0.170&	0.63&	0.28 – 1.45&	0.383\\
Center  [4]&	1.25&	0.46 – 3.43&	0.530&	1.67&	0.60 – 4.66&	0.187\\
Center  [6]&	0.11&	0.05 – 0.22&	0.034&	0.10&	0.05 – 0.21&	0.030\\
Race  * income $<40K$&	0.20&	0.14 – 0.30&	0.025&	0.16&	0.14 – 0.19&	0.014\\
Race  * income $>80K$&	0.47&	0.27 – 0.81&	0.450&	0.54&	0.54 – 0.55&	0.552\\
Time-to-surgery	&	&&&		1.21&	0.82 – 1.79&	0.029\\
Time-to-surgery$^2$&&&&				0.99&	0.57 – 1.75&	0.028\\
Race * Time-to-surgery&	&&&			1.17&	0.27 – 4.98&	0.416\\
Race *Time-to-surgery$^2$&	&&&			1.00&	0.13 – 7.45&	0.510\\

\end{tabular}
\end{table}

\begin{table}[!p]
\caption{
 Results of the mediation analysis in CanCORS dataset for stage II colon cancer patients (n=283). $TE_{60}$: racial difference in the $>60$ months survival probability, $SDE_{60}$: racial difference in the $>60$ months survival probability after the intervention on time to treatment distribution (shifting time to treatment distribution in blacks to match the one observed for the whites, i.e. reducing waiting times to surgery in the most disadvantaged group), and $PE$ the proportion of racial disparities eliminated by the intervention.}
\tabcolsep=4.25pt
\begin{center}
\begin{tabular}{@{}cccc@{}}
\bf{Method/Effects} & $TE_{60}$ & $SDE_{60}$ & $PE$ \\

Multistate & -0.29 (-0.51, -0.05)  & -0.26 (-0.47, -0.04)  & 8\%\\
exclude $T > S$ & -0.25 (-0.49, 0.00)  & -0.25 (-0.45, 0.00)  & 0\%\\
censor $T > S$ & -0.25 (-0.48, 0.00)  & -0.25 (-0.46, 0.00)  & 0\%\\
\end{tabular}
\end{center}
\end{table}

\newpage
\begin{figure}
\hspace*{-2cm}
\includegraphics[height=6.5in,width=6.5in]{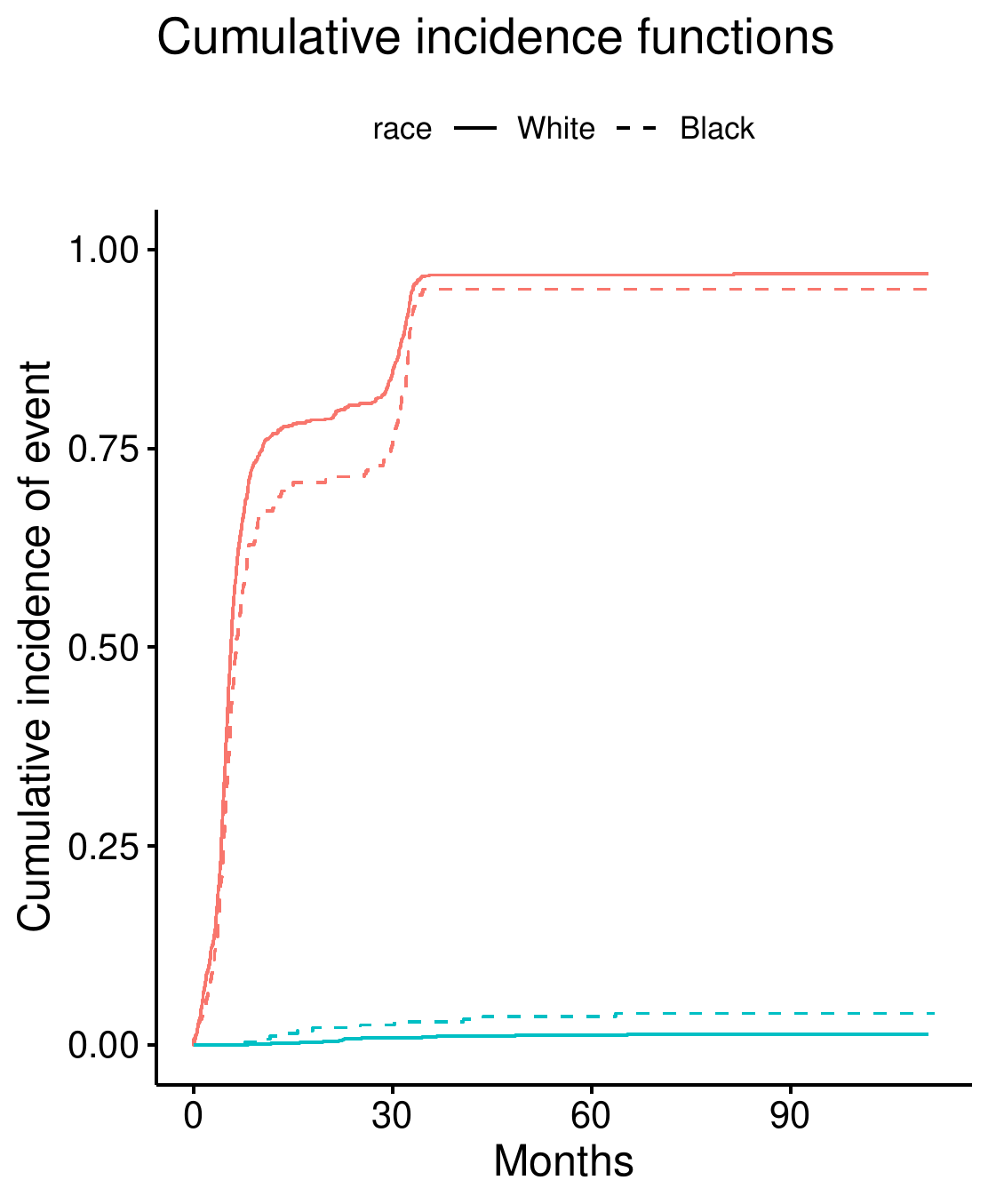}
\caption{Cumulative incidence curves stratified by racial-ethnic group for time-to-surgery (top curves) and time-to-death (bottom curves).}

\end{figure}

\vspace{-1in}
\begin{figure}
\hspace*{-2cm}
\includegraphics[height=7.5in,width=7.5in]{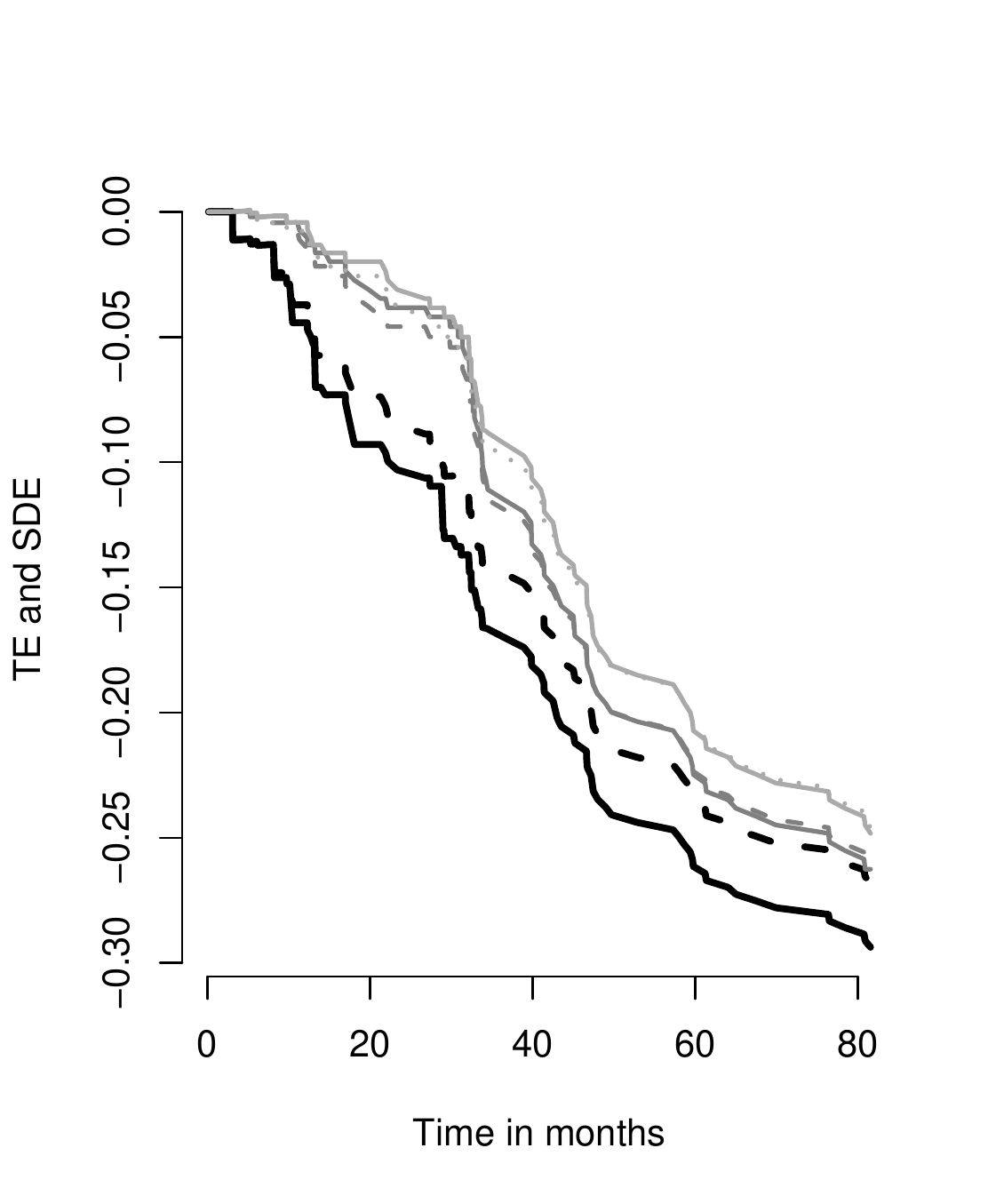}
\caption{Total effect ($TE$, solid line) and Stochastic direct effect ($SDE$, dashed line) on the survival probability difference scale estimated in the CanCORS data for subjects diagnosed at stage II under the three approaches: multistate model (black), exclude $T>S$ (dark gray) and censor $T>S$ (light gray).}

\end{figure}

 \end{document}